\documentclass[a4paper, 12pt]{article}
\usepackage{jheppub}
\usepackage{amssymb} 
\usepackage{amsmath}
\usepackage{mathtools}
\usepackage{amsfonts}    
\usepackage{dsfont}
\usepackage{young}
\usepackage[vcentermath]{youngtab}
\usepackage[mathscr]{euscript}
\usepackage{tensor}
\usepackage{xcolor}
\usepackage{graphicx}
\usepackage{fancyhdr}			
\usepackage{float}
\usepackage{subcaption}
\usepackage{comment}

\newcommand{\sltwof}{\mathfrak{sl}(2,\mathds{R})^{(1)}}
\newcommand{\sltwo}{\mathfrak{sl}(2,\mathds{R})}
\newcommand{\sutwof}{\mathfrak{su}(2)^{(1)}}
\newcommand{\sutwo}{\mathfrak{su}(2)}
\newcommand{\cJ}{\mathcal{J}}
\newcommand{\cK}{\mathcal{K}}

\newcommand{\be}{ \begin{equation}}
\newcommand{\ee}{\end{equation}}

\renewcommand{\|}{|\!|}

\begin{titlepage}
\date{\hfill \today}
\title{D-Branes in $\textbf{AdS}_3\times \textbf{S}^3 \times \textbf{S}^3 \times \textbf{S}^1$}
\end{titlepage}
\makeatother

\author[a]{Giorgio Belleri}
\author[a,b]{and Matthias R.\ Gaberdiel}
\affiliation[a]{Institut f\"ur Theoretische Physik,
ETH Z\"urich,\\
Wolfgang-Pauli-Strasse 27,
8093 Z\"urich, Switzerland}
\affiliation[b]{Kavli Institute for Theoretical Sciences, University of Chinese Academy of Sciences,\\
Beijing 100190, China}
\emailAdd{gbelleri@ethz.ch}
\emailAdd{gaberdiel@itp.phys.ethz.ch}

\abstract{String theory on ${\rm AdS}_3\times {\rm S}^3 \times {\rm S}^3 \times {\rm S}^1$ with two units of NS flux through each of the two $3$-spheres (and one unit of NS flux supporting the ${\rm AdS}_3$ factor) was recently argued to be exactly dual to the symmetric orbifold of eight free fermions and two bosons. This setup is interesting since it allows for a simple NS-R description. In this paper we study the spherical D-branes of this theory, and identify them with branes in the dual symmetric orbifold theory. We also comment briefly on the ${\rm AdS}_2$ branes.}

\begin{document}
\maketitle

\section{Introduction}

Much progress has recently been made with identifying exact AdS/CFT dual pairs. In particular, for the $3$-dimensional AdS case, it was shown that string theory on ${\rm AdS}_3 \times {\rm S}^3 \times \mathbb{T}^4$ with one unit ($k=1$) of NS-NS flux through both the ${\rm S}^3$ and ${\rm AdS}_3$ is exactly dual to the symmetric orbifold of $\mathbb{T}^4$ \cite{Eberhardt:2018ouy,Eberhardt:2019ywk}, see also \cite{Dei:2020zui,Eberhardt:2020akk,Knighton:2020kuh,McStay:2023thk,Dei:2023ivl,Knighton:2023mhq} for subsequent work. While much has been understood about this duality, one slightly complicating issue is that the worldsheet theory cannot be directly formulated in the NS-R formalism since the $\mathfrak{su}(2)$ factor, describing the propagation of strings on the ${\rm S}^3$, is at (supersymmetric) level $k=1$, and hence involves a bosonic $\mathfrak{su}(2)$ algebra at level $k=-1$ after decoupling the fermions. This has necessitated working with the somewhat more complicated hybrid formalism of \cite{Berkovits:1999im}.

Recently, it was noted in \cite{Gaberdiel:2024dva} that string theory on ${\rm AdS}_3 \times {\rm S}^3 \times {\rm S}^3 \times {\rm S}^1$ with two units ($k=2$) of NS-NS flux through both ${\rm S}^3$ (and one unit of NS-NS flux through the ${\rm AdS}_3$) may be exactly dual to the symmetric orbifold of eight free fermions and two bosons. This idea goes already back to the early days of the tensionless string duality \cite{Gaberdiel:2018rqv,Giribet:2018ada}, but further arguments became available following the more recent developments of \cite{Dei:2023ivl,Knighton:2023mhq,Knighton:2024qxd,Sriprachyakul:2024gyl,Sriprachyakul:2024xih,Hikida:2023jyc,Knighton:2024pqh}. In addition, it was recently shown in \cite{Eberhardt:2025sbi} that the worldsheet theory of \cite{Gaberdiel:2024dva} is consistent without the inclusion of any additional states.\footnote{This invalidates the criticism of \cite{Chakraborty:2025nlb}.} One attractive feature of the proposal of \cite{Gaberdiel:2024dva} is that the worldsheet theory can be formulated in the easier NS-R description, and this should allow one to study and understand it in quite some detail. In this paper we take one small step in this direction by analysing the so-called `spherical' D-branes of ${\rm AdS}_3$ in this set-up. For the case of 
${\rm AdS}_3 \times {\rm S}^3 \times \mathbb{T}^4$ this was already done using the hybrid description in \cite{Gaberdiel:2021kkp}, see also \cite{Belin:2021nck,Martinec:2022ofs,Gutperle:2024vyp,Knighton:2024noc,Gutperle:2024rvo,Harris:2025wak} for recent discussions of boundary states and defects in ${\rm AdS}_3/{\rm CFT}_2$; here we repeat this analysis in the NS-R formalism for the ${\rm AdS}_3 \times {\rm S}^3 \times {\rm S}^3 \times {\rm S}^1$ background. We also briefly comment on the ${\rm AdS}_2$ branes in Section~\ref{sec:AdS2}; they were recently identified with certain interfaces in the symmetric orbifold in \cite{Harris:2025wak}.

\medskip

The paper is organised as follows. In Section~\ref{sec:worldsheet} we review our conventions for the worldsheet theory. Section~\ref{Sec:gluing} describes the boundary conditions that we shall impose on the worldsheet fields, as well as the relevant boundary states and their cylinder diagrams. In Section~\ref{Sec:Orbifold} we turn to the dual symmetric orbifold and explain how the relevant boundary states can be constructed for that theory. We also explain there how the corresponding cylinder amplitudes reproduce those of the worldsheet calculation. Section~\ref{sec:AdS2} briefly comments on the ${\rm AdS}_2$ branes, and we close with some conclusions in Section~\ref{Sec:conclusions}. Some of the technical material has been delegated to two appendices.

\section{The worldsheet theory}\label{sec:worldsheet}

Let us start by explaining our conventions for the description of the worldsheet theory. For pure NS-NS flux we can describe the background in terms of a WZW model following \cite{Maldacena:2000hw}
\begin{equation}
\label{WZW}
\mathfrak{sl}(2,\mathds{R})^{(1)}_k \oplus \mathfrak{su}(2)^{(1)}_{k_+} \oplus \mathfrak{su}(2)^{(1)}_{k_-}  \oplus \mathfrak{u}(1)^{(1)} \ . 
\end{equation}
Criticality of the worldsheet theory requires that \cite{Elitzur:1998mm}
\be
\frac{1}{k} = \frac{1}{k_+} + \frac{1}{k_-} \ , 
\ee
and we will mainly concentrate on the case where 
\be
k_+ = k_- = 2 \ , \qquad k=1 \ . 
\ee
For $\sltwof_k$ we shall work with the conventions \cite{Maldacena:2000hw,Ferreira:2017pgt}
\begin{alignat}{5}
\bigl[J^{+}_{m},J^{-}_{n}\bigr] 
={}&
 -2J^{3}_{m+n} + km\delta_{m,-n}&
 & \quad &
 \bigl[J^{3}_{m},J^{\pm}_{n}\bigr] 
={}&
 \pm J^{\pm}_{m+n}&
  & \quad &
 \bigl[J^{3}_{m},J^{3}_{n}\bigr] 
={}&
 -\tfrac{k}{2}m\delta_{m,-n}
 \nonumber\\
 \bigl[J^{\pm}_{m},\psi^{3}_{r}\bigr]
={}&
 \mp\psi^{\pm}_{m+r} &
 & \quad &
 \bigl[J^{3}_{m},\psi^{\pm}_{r}\bigr]
={}&
 \pm\psi^{\pm}_{m+r}&
  &\quad &
 \bigl[J^{\pm}_{m},\psi^{\mp}_{r}\bigr]
 ={}&
 \mp 2\psi^{3}_{m+r}
 \nonumber \\
 \bigl\{\psi^{+}_{r},\psi^{-}_{s}\bigr\}
={}&
 k\delta_{r,-s}&
 & \quad &
 \bigl\{\psi^{3}_{r},\psi^{3}_{s}\bigr\}
={}&
 -\tfrac{k}{2}\delta_{r,-s}\ .  &
\end{alignat}
We will also denote the decoupled currents by\footnote{Recall that the dual Coxeter number of $\sltwo$ is $h^\vee = -2$, while that of $\sutwo$ is $h^\vee=+2$.} 
\be\label{Jcal}
\cJ^{3} 
={} J^{3} +\tfrac{1}{k}\bigl(\psi^{-}\psi^{+}\bigr) \ , \qquad 
\cJ^{\pm} 
={} 
J^{\pm} \pm \tfrac{2}{k}\bigl(\psi^{3}\psi^{\pm}\bigr) \ . 
\ee
By construction, they then commute with the fermions, $[\cJ^{a}_{n},\psi^{b}_{r}]=0$, and satisfy the same algebra as the $J^{a}$ at level $\kappa = k+2$. For each ${\rm S}^3$ factor we have instead an $\sutwof_{k}$ algebra, (with $k=k_\pm$)  for which our conventions are
\begin{alignat}{5}
\bigl[K^{+}_{m},K^{-}_{n}\bigr] 
={}&
 2K^{3}_{m+n} + k  m\delta_{m,-n}&
 &\quad &
 \bigl[K^{3 }_{m},K^{\pm}_{n}\bigr] 
={}&
 \pm K^{\pm}_{m+n}&
 &\quad &
 \bigl[K^{3}_{m},K^{3}_{n}\bigr] 
={}&
 \tfrac{k}{2}m\delta_{m,-n}
 \nonumber\\
 \bigl[K^{\pm}_{m},\chi^{3}_{r}\bigr]
 ={}&
 \mp\chi^{\pm}_{m+r}&
 &\quad &
 \bigl[K^{3(\pm)}_{m},\chi^{\pm}_{r}\bigr]
 ={}&
 \pm\chi^{\pm}_{m+r}&
 &\quad &
 \bigl[K^{\pm}_{m},\chi^{\mp}_{r}\bigr]
 ={}&
 \pm 2\chi^{3}_{m+r}
\nonumber  \\
 \bigl\{\chi^{+}_{r},\chi^{-}_{s}\bigr\}
 ={}&
 k\delta_{r,-s}&
 &\quad &
 \bigl\{\chi^{3}_{r},\chi^{3}_{s}\bigr\}
 ={}&
 \tfrac{k}{2}\delta_{r,-s} \ .&
\end{alignat}
As for the case of $\sltwof$, we can decouple the bosons from the fermions
by defining 
\begin{equation}\label{decoupled K currents 1}
\mathcal{K}^{3} =  K^{3} -\tfrac{1}{k}\left(\chi^{+}\chi^{-}\right) \ , \qquad \mathcal{K}^{\pm} =  K^{\pm} \mp \tfrac{2}{k}\left(\chi^{3}\chi^{\pm}\right)\ ,
% \label{decoupled K currents 2}
\end{equation}
so that $\bigl[\cK^{a}_{m},\chi^{b}_{n}\bigr] =0$. The decoupled currents satisfy again the same algebra as the $K^{a}$, but at level $\kappa = k-2$ instead. For the case at hand, $k_\pm=2$, and hence the decoupled bosons will have vanishing level $\kappa_{\pm} = k_\pm -2 = 0$. In particular, these currents are therefore null and do not contribute to the physical spectrum. 

Following \cite{Gaberdiel:2024dva} the spectrum of the (decoupled) $\mathfrak{sl}(2,\mathds{R})_3$ worldsheet theory consists of the continuous representations $\mathscr{F}_{\lambda,s}$ with $j=\frac{1}{2} + is$, where $s\in\mathbb{R}$ and $\lambda$ denotes the eigenvalue of ${\cal J}^3_0$ mod integers. Furthermore, as in \cite{Maldacena:2000hw}, see also \cite{Henningson:1991jc}, we  need to include the spectrally flowed images of these representations, where spectral flow acts on the (coupled) $\sltwof_1$ algebra as 
\begin{equation}\label{J3flow}
\begin{array}{rclrcl}
\sigma^w(J^{3}_n) & =& J^{3}_n+ \frac{kw}{2}\delta_{n,0} = J^{3}_n+ \frac{w}{2}\delta_{n,0} \ ,\qquad & \sigma^w(J^{\pm}_n) & = & J^{\pm}_{n\mp w}\ ,
\end{array}
\end{equation}
since $k=1$. For the worldsheet Virasoro algebra this then induces the action 
\be\label{Lflow}
\sigma^w(L_n)  =  L_n-w J^3_n - k \tfrac{w^2}{4}\, \delta_{n,0}  = L_n-w J^3_n -  \tfrac{w^2}{4}\, \delta_{n,0}   \ ,
\ee
where we have again set $k=1$. Without loss of generality, we take spectral flow only to act on the $\mathfrak{sl}(2,\mathds{R})$ factor, but not on the two $\sutwo$ algebras,\footnote{For $\sutwo$ spectral flow only rearranges the familiar highest weight representations, and therefore does not modify the worldsheet spectrum. Also, as we explained above, the decoupled bosonic generators of $\sutwo$ at level $\kappa_\pm=0$ vanish, and the two $\mathfrak{su}(2)$ factors therefore only contribute six fermions.} and the full worldsheet spectrum is then of the form 
\begin{equation}
\label{wsspec}
{\cal H} =  \Bigl[ \sum_{w\in\mathbb{Z}} \int_0^1d \lambda \int_{\mathds{R}} ds\, \sigma^w \bigl({\cal H}_{j=\frac{1}{2}+is,\lambda} \bigr) \otimes \sigma^w \bigl(\bar{\cal H}_{j=\frac{1}{2}+is,\lambda} \bigr) \Bigr] \otimes {\cal H}^{S^1} \otimes{\cal H}_{\rm fermions} \ .
\end{equation}

\section{Gluing conditions for spherical branes}
\label{Sec:gluing}

The general geometry of D-branes in ${\rm AdS}_3$ was studied in \cite{Bachas:2000fr}, and subsequently various aspects of these branes have been analysed in  \cite{Giveon:2001uq,Petropoulos:2001qu,Lee:2001xe,Hikida:2001yi,Rajaraman:2001cr,Lee:2001gh,Ponsot:2001gt}. As in \cite{Gaberdiel:2021kkp} we shall mainly consider the so-called spherical branes\footnote{The name `spherical' goes back to \cite{Ponsot:2001gt}, and refers to the fact that the branes describe $2$-spheres in the corresponding Euclidean background. With respect to ${\rm AdS}_3$, these branes describe an instantonic hyperbolic $2$-plane.} that are characterised by the boundary condition at $z=\bar{z}$
\begin{align}
J^{\, 3}(z)=-\bar{J}^{\, 3}(\bar{z})\ , \qquad 
 J^{\, \pm}(z)=\bar{J}^{\, \mp}(\bar{z})\ ,  \label{eq:GluJIdp}
\end{align}
where the right-movers are denoted by a bar. By supersymmetry, the corresponding fermionic fields then satisfy
\begin{align}
\psi^{\, 3}(z)=-\varepsilon \,  \bar{\psi}^{\, 3}(\bar{z})\ , \qquad 
 \psi^{\, \pm}(z)=\varepsilon \,  \bar{\psi}^{\, \mp}(\bar{z}) \label{eq:GluJIdpf} \ ,
\end{align}
where $\varepsilon=\pm$ is the worldsheet spin structure. 

The boundary conditions along the two ${\rm S}^3$'s and the ${\rm S}^1$ will not have much impact on our analysis. For the two ${\rm S}^3$ factors, we only have the six fermions $\chi^{\, a (\pm)}$, where $a\in\{\pm ,3\}$ (since the decoupled bosonic generators vanish), and we impose the same boundary conditions on the fermions as in  (\ref{eq:GluJIdpf}),  i.e.\footnote{Since the natural light-cone direction is a linear combination of $J^3$ and $K^{3(\pm)}$, it is natural to impose the same gluing condition for $K^{3(\pm)}$ as for $J^3$, and similarly for the fermions.\label{foot5}}
\begin{align}
\chi^{\, 3(\pm)}(z)=-\varepsilon \,  \bar{\chi}^{\, 3(\pm)}(\bar{z})\ , \quad 
 \chi^{\, +(\pm)}(z)=\varepsilon \,  \bar{\chi}^{\, -(\pm)}(\bar{z}) \ ,  \quad 
 \chi^{\, -(\pm)}(z)=\varepsilon \,  \bar{\chi}^{\, +(\pm)}(\bar{z}) \ . 
  \label{eq:GluJIdpS3}
\end{align}
Finally, we take the boundary condition along the ${\rm S}^1$ to be of Neumann type.

In eq.~(\ref{eq:GluJIdp}) we have formulated the boundary conditions for $\mathfrak{sl}(2,\mathds{R})$ in terms of the coupled currents. However, the fermionic gluing conditions (\ref{eq:GluJIdpf}) imply that 
\be
\psi^3(z)\, \psi^\pm (z) = - \bar{\psi}^3(\bar{z})\, \bar{\psi}^{\mp} (\bar{z})  \ , \quad \hbox{and} \quad 
\psi^-(z)\, \psi^+(z) = -  \bar{\psi}^{-}(\bar{z}) \, \bar{\psi}^{+}(\bar{z})  \ , 
\ee
and thus the same boundary conditions apply then also to the decoupled currents of eq.~(\ref{Jcal}). Similarly, the boundary conditions for the fermions in (\ref{eq:GluJIdp}) imply that the corresponding coupled $\mathfrak{su}(2)$ generators satisfy 
\be
K^{\, 3 (\pm)}(z)=- \bar{K}^{\, 3 (\pm)}(\bar{z})\ , \quad 
K^{\, + (\pm)}(z)= \bar{K}^{\, - (\pm)}(\bar{z})\ , \quad 
K^{\, - (\pm)}(z)= \bar{K}^{\, + (\pm)}(\bar{z})\ .
\ee

\subsection{Ishibashi states and gluing conditions}

For the analysis of boundary conditions it is often convenient to use the description in terms of boundary states that `imitate' the boundary conditions from the corresponding closed string perspective, see e.g.\ \cite{Gaberdiel:2002my} for a brief introduction. The boundary conditions from above then translate into gluing conditions for the corresponding boundary states, and for the case at hand, these gluing conditions take the form 
\begin{subequations}
\label{eq:IshibashiSpherical}
\begin{align}
    (J_n^3-\bar{J}_{-n}^3)|w,\lambda,s,\varepsilon\rangle\!\rangle_{\mathrm{S}} &=0\ , \label{J3cond}\\
    (J_n^\pm+\bar{J}_{-n}^\mp)|w,\lambda,s,\varepsilon\rangle\!\rangle_{\mathrm{S}} &=0\ ,\\
    (K_n^{\, 3 (\pm)}-\bar{K}_{-n}^{\, 3 (\pm)})|w,\lambda,s,\varepsilon\rangle\!\rangle_{\mathrm{S}} &=0\ ,\\
    (K_n^{\, \pm (\pm)}+\bar{K}_{-n}^{\, \mp (\pm)})|w,\lambda,s,\varepsilon\rangle\!\rangle_{\mathrm{S}} &=0\ ,\\
    (\psi_r^3-i\varepsilon \,  \bar{\psi}_{-r}^3)|w,\lambda,s,\varepsilon\rangle\!\rangle_{\mathrm{S}} &=0\ , \label{eq:gc_ferm}\\
    (\psi_r^\pm+i\varepsilon \,  \bar{\psi}_{-r}^\mp)|w,\lambda,s,\varepsilon\rangle\!\rangle_{\mathrm{S}} &=0\ , \label{eq:gc_ferm2}\\
    (\chi_r^{\, 3 (\pm)}-i\varepsilon \,  \bar{\chi}_{-r}^{\, 3 (\pm)})|w,\lambda,s,\varepsilon\rangle\!\rangle_{\mathrm{S}} &=0\ ,\\
    (\chi_r^{\, \pm (\pm)}+i\varepsilon \,  \bar{\chi}_{-r}^{\, \mp (\pm)})|w,\lambda,s,\varepsilon\rangle\!\rangle_{\mathrm{S}} &=0\ ,\\
    (\alpha_n^1+\bar{\alpha}_{-n}^1)|w,\lambda,s,\varepsilon\rangle\!\rangle_{\mathrm{S}} &=0\ ,  \label{eq:gc_circleb} \\
    (\psi_r^1+i\varepsilon \,  \bar{\psi}_{-r}^1)|w,\lambda,s,\varepsilon\rangle\!\rangle_{\mathrm{S}} &=0\ , 
    \label{eq:gc_circle}\
\end{align}
\end{subequations}
where the label `S' indicates that we are considering here the `spherical' branes. 

The condition (\ref{J3cond}) with $n=0$ implies that the left- and right-moving $J^3_0$ eigenvalues must be equal. Since in the worldsheet spectrum of eq.~(\ref{wsspec}) we have $\lambda=\bar{\lambda}$, this is possible for all $\mathfrak{sl}(2,\mathds{R})$ representations that appear in eq.~(\ref{wsspec}). (This is in contrast to the so-called ${\rm AdS}_2$ branes for which the gluing condition is $J^3_0 + \bar{J}^3_0=0$, and only few sectors can support Ishibashi states, see e.g.\ \cite{Gaberdiel:2021kkp} and the discussion in Section~\ref{sec:AdS2}.) The other feature that distinguishes the spherical branes is that there exists an Ishibashi state in each $w$-spectrally flowed sector. Indeed, the relevant Ishibashi state is simply obtained from the unflowed Ishibashi state via 
\begin{align}
 |w,\lambda,s,\varepsilon\rangle\!\rangle_{\mathrm{S}}=\big[|0,\lambda,s,\varepsilon\rangle\!\rangle_{\mathrm{S}}\big]^w  \ ,
\end{align}
and it exists (i.e.\ is non-trivial) for all $w\in\mathbb{Z}$, $\lambda\in [0,1)$ and $s\in\mathbb{R}$.

\subsection{Cylinder diagrams for $\mathfrak{sl}(2,\mathds{R})_3$}

In order to determine the actual boundary states (i.e.\ the specific linear combinations of the Ishibashi states that satisfy the Cardy condition) one needs to determine their cylinder diagrams. For the individual Ishibashi states associated to the decoupled $\mathfrak{sl}(2,\mathds{R})_3$ factor, the cylinder diagrams take the form\footnote{Since we are only considering the decoupled  $\mathfrak{sl}(2,\mathds{R})_3$ factor, the spin structure does not enter the definition of the Ishibashi or boundary states.}
\begin{align}
&    _{\mathrm{S}}\langle\!\langle w',\lambda',s'| \hat{q}^{\frac{1}{2}(L_0+\bar{L}_0-\frac{c}{12})}\hat{x}^{\frac{1}{2}(J_0^3+\bar{J}_0^3)}|w,\lambda,s\rangle\!\rangle_{\mathrm{S}} & \nonumber \\
& \qquad \qquad = 
        \delta_{w,w'}\, \delta(\lambda'-\lambda)\,\delta(s'-s) \, {\mathrm{ch}}[\sigma^{w}(\mathscr{F}_{\lambda,s})](\hat{t};\hat{\tau}) \ , 
\end{align}
where we have denoted the closed string parameters by a hat, 
\begin{equation}
\hat{q}=e^{2\pi i \hat{\tau}}   \ , \qquad 
\hat{x}=e^{2\pi i \hat{t}} \ ,
\end{equation}
and $\hat{t}$ is the `modular' parameter of the dual spacetime CFT, i.e.\ the chemical potential for $J^3_0$. 
The relevant characters are explicitly, see e.g.\ \cite{Gaberdiel:2024dva}
\begin{align}
\label{eq:characterSL}
{\mathrm{ch}}[\sigma^{w}(\mathscr{F}_{\lambda,s})]({t};{\tau})& = 
\eta({\tau})^{-3}e^{2\pi i{\tau}(s^2 +\frac{3 w^2}{4})}\sum_{m\in\mathbb{Z}}e^{2\pi im(\lambda+\frac{3w}{2})}\delta({t}-w{\tau}-m) \\
& = \eta({\tau})^{-3}e^{2\pi i{\tau}(s^2 -\frac{3 w^2}{4} - w\lambda)}e^{2\pi i t (\lambda+\frac{3w}{2})}\sum_{m\in\mathbb{Z}}\delta({t}-w{\tau}-m) \ . \label{eq:ch_w}
\end{align}
Their $S$-modular transformation can be obtained as in \cite{Baron:2010vf}, see also \cite{Eberhardt:2018ouy,Gaberdiel:2024dva}, and one finds\footnote{For the theory at hand, $s\in\mathbb{R}$, and thus we just get an exponential in $s$ instead of the cosine of say  \cite[eq.~(3.10)]{Baron:2010vf}.}
\begin{align}
& {\mathrm{ch}}[\sigma^{w}(\mathscr{F}_{\lambda,s})](\tfrac{{t}}{{\tau}};-\tfrac{1}{{\tau}}) \\
& \quad =e^{-\frac{3}{2}\pi i\frac{{t}^2}{{\tau}}}\sum_{w'=-\infty}^{\infty} 
\int_{ -\infty}^{+\infty}ds' \int_0^1 d\lambda'~
{\cal S}_{s\, \lambda\, w}{}^{s'\lambda'w'} \, {\mathrm{ch}}[\sigma^{w}(\mathscr{F}_{\lambda,s})]({t};{\tau}) \ ,\nonumber
\end{align}
with the $S$-matrix 
\begin{equation}
{\cal S}_{s\,\lambda\,w}{}^{s'\lambda'w'}= i \sqrt{2} \, \frac{|\tau|}{\tau}\, e^{4\pi  iss'}\, e^{2\pi i \left(w\lambda'+w'\lambda+\frac{3}{2}
  ww'\right)}\ .
   \label{Smat}
\end{equation}
As in \cite{Gaberdiel:2021kkp}, we can now make an ansatz for the $\mathfrak{sl}(2,\mathds{R})_3$ boundary states as 
\begin{align}
\| W,\Lambda,S\rangle\!\rangle_\mathrm{S} = 2^{\frac{1}{4}}\sum_{w\in \mathbb{Z}}\int_0^1 d\lambda\int_{-\infty}^\infty ds\, e^{2\pi i [w\Lambda+\lambda W+sS]}\, |w,\lambda,s\rangle\!\rangle_\mathrm{S} \ ,\label{eq:BSspherical}
\end{align}
where $W\in\mathbb{Z}$, while $\Lambda\in[0,1)$ and $S\in\mathbb{R}$. Their overlaps then satisfy 
\begin{align}
\hat{Z}&_{(W_1,\Lambda_1,S_1)|(W_2,\Lambda_2,S_2)}(\hat{t};\hat{\tau})\equiv\,_{\text{S}}\langle\!\langle W_2,\Lambda_2,S_2\| \hat{q}^{\frac{1}{2}(L_0+\bar{L}_0-\frac{c}{12})}\hat{x}^{\frac{1}{2}(J_0^3+\bar{J}_0^3)}\|W_1, \Lambda_1,S_1\rangle\!\rangle_{\text{S}} \nonumber \\
&=\sqrt{2}\sum_{w\in\mathbb{Z}}\int_0^1 d\lambda \,
\int_{-\infty}^\infty ds \, e^{2\pi i w(\Lambda_1-\Lambda_2)} \,  
e^{2\pi i \lambda(W_1-W_2)} \, e^{2\pi i s(S_1-S_2)} \, {\text{ch}}[\sigma^w(\mathscr{F}_{\lambda,s})](\hat{t};\hat{\tau})\ . \label{3.14b}
\end{align}
By performing the modular $S$-transformation to the open-string channel, we then obtain
\begin{align}
\hat{Z}_{(W_1,\Lambda_1, S_1)|(W_2,\Lambda_2, S_2)}(\hat{t};\hat{\tau})
	&= e^{-\frac{3 \pi i \tau}{2t^2}}\, {\mathrm{ch}}[\sigma^{W_2-W_1}(\mathscr{F}_{\Lambda_2-\Lambda_1-\frac{3}{2}(W_2-W_1),\frac{1}{2}(S_2-S_1)})]\left(-\tfrac{\tau}{t};\tau\right)\ ,\label{eq:SphericalOpen}
\end{align}
where we have defined the open-string variables $t$ and $\tau$ via 
\begin{equation}\label{openstring}
    \hat{t}=- \frac{1}{t}\ , \qquad     \hat{\tau}=-\frac{1}{\tau}\ , 
\end{equation}
and used the identities 
\be
\sum_{w\in\mathbb{Z}} e^{2\pi i w a}  = \sum_{m\in\mathbb{Z}}\, \delta(a-m) \ , \quad 
\int_0^1 d\lambda \, e^{2\pi i \lambda \, \Delta W}  = \delta_{\Delta W,0} \ , \quad 
\int_{-\infty}^{\infty} ds \, e^{2\pi i s \Delta S}  = \delta(\Delta S) \ ,
\ee
where $\Delta W\in\mathbb{Z}$. 
The overall factor of $e^{-\frac{3 \pi i \tau}{2t^2}}$ is the usual factor that arises in the transformation of Jacobi forms, and the resulting open string spectrum is therefore consistent --- it is just equal to the representation whose character appears on the right-hand-side. For example, for $W_1=W_2$, $\Lambda_1=\Lambda_2$  and $S_1=S_2$, the spectral flow in the open string is trivial, and the open string spectrum consists of the vacuum representation $\mathscr{F}_{0,0}$.

\subsection{The full boundary states on the worldsheet }\label{sec:cylinder}

In the previous section we have only discussed the boundary states for $\mathfrak{sl}(2,\mathds{R})_3$, but the full worldsheet theory also consists of the ${\rm S}^1$ factor, as well as the free fermions. The construction of boundary states for ${\rm S}^1$ is briefly reviewed in Appendix~\ref{appCircle}, and we denote the Neumann brane with Wilson line $\theta$ by $\|\theta\rangle\!\rangle$, see eq.~(\ref{Dirbrane}). For the free fermions, the boundary states are discussed in Appendix~\ref{appFermions}. The full boundary state of our worldsheet theory is then the product of these different boundary components, and we denote it by 
\be
\label{eq:fullBS}
\| W,\Lambda,S,\theta,\varepsilon;\hat{t}\,\rangle\!\rangle_\mathrm{S} \equiv 
\| W,\Lambda,S;\hat{t}\,\rangle\!\rangle_\mathrm{S} \otimes \|\theta\rangle\!\rangle \otimes \| {\rm F}, \varepsilon \rangle\!\rangle \ , 
\ee
where ${\rm F}$ denotes the boundary state for the ten fermions. The worldsheet theory needs to be GSO-projected, and in the $w$-spectrally flowed sector the ground state has GSO-parity $\frac{w+1}{2}$, i.e.\ it is GSO-invariant if $w$ is odd, see e.g.\ \cite[eq.~(5.7)]{Ferreira:2017pgt}. As a consequence, the GSO-invariant Ishibashi (or boundary) states in the NS-NS sector are of the form 
\begin{equation}
\label{eq:GSOstateNS}
\| {\rm NS}\rangle\!\rangle_{\text{S}} = \begin{cases} 
  \frac{1}{{2}}\bigl[| {\rm NS}, +\rangle\!\rangle_{\text{S}} + | {\rm NS}, -\rangle\!\rangle_{\text{S}} \bigr] \ , 
  & \text{if }w=\text{ odd}\\ 
 \frac{1}{{2}}\bigl[| {\rm NS}, +\rangle\!\rangle_{\text{S}} - | {\rm NS}, -\rangle\!\rangle_{\text{S}}\bigr] \ , 
  & \text{if }w=\text{ even} \ ,
    \end{cases}   
\end{equation}
where we have suppressed the other labels of the boundary states. In the R-R sector the situation is similar
\begin{equation}
\label{eq:GSOstateR}
\| {\rm R}\rangle\!\rangle_{\text{S}} = \begin{cases} 
  \frac{1}{{2}}\bigl[| {\rm R}, +\rangle\!\rangle_{\text{S}} + | {\rm R}, -\rangle\!\rangle_{\text{S}} \bigr]\ , 
  & \text{if }w=\text{ odd}\\ 
 \frac{1}{{2}}\bigl[| {\rm R}, +\rangle\!\rangle_{\text{S}} - | {\rm R}, -\rangle\!\rangle_{\text{S}} \bigr]\ , 
  & \text{if }w=\text{ even} \ .
    \end{cases}   
\end{equation}
For the fermions, the gluing conditions (\ref{eq:IshibashiSpherical}) impose, for each choice of $\varepsilon$,  four gluing conditions with $+ i \varepsilon$, and six with $-i \varepsilon$; the relevant R-R boundary states are therefore compatible with a type IIB GSO projection, and this is what we shall have in mind. For the worldsheet we shall work with the combinations
\be
\label{eq:GSOstate}
\| {\rm F},\pm\rangle\!\rangle = \| {\rm NS}\rangle\!\rangle_{\text{S}} \pm \| {\rm R}\rangle\!\rangle_{\text{S}}  \  , 
\ee
and their overlaps equal 
\begin{align}
\label{eq:GSOstateOverlap}
& \langle\!\langle{\rm F},\pm\| \hat{q}^{\frac{1}{2}(L_0+\bar{L}_0-\frac{c}{12})} \| {\rm F},\pm\rangle\!\rangle \\
& \qquad = 
\frac{1}{2} \Bigl( \hat{\chi}^+_{{\rm NSNS}}(\hat{\tau})+ (-1)^{w+1}\hat{\chi}^-_{{\rm NSNS}}(\hat{\tau})+\hat{\chi}^+_{{\rm RR}}(\hat{\tau})+ (-1)^{w+1}\hat{\chi}^-_{{\rm RR}}(\hat{\tau})\Bigr) \ , 
\end{align} 
where the overlaps are given in eq.~(\ref{eq:NSRoverlaps}). Notice that the RR overlap with opposite spin structure (i.e.\ $\hat{\chi}^-_{\rm RR}$) vanishes.
\smallskip

\noindent We are now ready to calculate the overlap for the full boundary states
\be
\label{eq:full_overlap}
\hat{Z}(\hat{t};\hat{\tau})  = {}_{\mathrm{S}}\langle\!\langle W, \Lambda, S \| \otimes \langle\!\langle \theta \| \otimes \langle\!\langle \mathrm{F}, \varepsilon \| \hat{q}^{\frac{1}{2}(L_0 + \bar{L}_0-\frac{c}{12})}\hat{x}^{\frac{1}{2}(J^3_0 + \bar{J}^3_0)}\| W,\Lambda,S\,\rangle\!\rangle_\mathrm{S} \otimes \|\theta\rangle\!\rangle \otimes \| {\rm F}, \varepsilon \rangle\!\rangle \ .
\ee
To do so, we will work with the decoupled currents and factorise the Hamiltonian and $J^3_0$ into their bosonic and fermionic parts, e.g.\ $J^3_0 = \mathcal{J}^3_0+J^{3,(\mathrm{f})}_0$. The bosonic $\mathfrak{sl}(2,\mathds{R})_3$ part was already studied in the previous section,\footnote{The contribution coming from ${\rm S}^1$ is standard, and is worked out in Appendix~\ref{appCircle}.} and we will therefore focus on the contribution from the fermions, i.e.
\begin{equation}
\label{eq:ferm_x}
    \langle\!\langle \mathrm{F}, \varepsilon \| \hat{q}^{\frac{1}{2}(L_0^{\rm(f)} + \bar{L}_0^{\rm(f)}-\frac{c}{12})}\hat{x}^{\frac{1}{2}(J^{3,{\rm(f)}}_0 + \bar{J}^{3,{\rm(f)}}_0)} \| {\rm F}, \varepsilon \rangle\!\rangle \ .
\end{equation}
It is convenient to also spectrally flow the $\mathfrak{sl}(2,\mathds{R})$ fermions --- in fact, this is implicit in the above convention for the GSO-projection --- and then the associated stress-energy tensor and current transform as 
\begin{subequations}
\begin{align}
        & J_0^{3, {\rm (f)}} \rightarrow J_0^{3, {\rm (f)}} - w \ , \\ 
        & L_0^{\rm (f)} \rightarrow L_0^{\rm (f)} - w J_0^{3, {\rm (f)}} + \frac{w^2}{2} \ , 
\end{align}
\end{subequations}
i.e.\ they behave as a `level' $k=-2$ $\mathfrak{sl}(2,\mathds{R})$ algebra, c.f.\ eq.~(\ref{J3flow}) and (\ref{Lflow}). For example, in the NS sector with matching spin structures, the overlap for the $10$ fermions equals 
\be
\,_{\text{S}}\langle\!\langle{\rm NS},\pm\| \hat{q}^{\frac{1}{2}(L_0+\bar{L}_0-\frac{c}{12})} \hat{x}^{\frac{1}{2}(J^{3,\mathrm{(f)}}_0+\bar{J}^{3,\mathrm{(f)}}_0)} \| {\rm NS},\pm\rangle\!\rangle_{\text{S}}  = \hat{q}^{\frac{w^2}{2}}\hat{x}^{-w}\, \frac{\vartheta_3(\hat{t}-w\hat{\tau};\hat{\tau})\,\vartheta_3(\hat{\tau})^4}{\eta(\hat{\tau})^5}  \ ,
\ee
where the two fermions charged under $J^{3,\mathrm{(f)}}_0$ give rise to the $\vartheta_3(\hat{t}-w\hat{\tau};\hat{\tau})$ factor. The contributions in the other sectors work analogously.

Combining with the bosons, and using the form of the character of eq.~(\ref{eq:ch_w}), the full boundary states of eq.~(\ref{eq:fullBS}) then give rise to the overlap, see eq.~(\ref{eq:full_overlap})
\begin{align*}
\label{eq:closed_ampl}
\hat{Z}(\hat{t};\hat{\tau}) =
    & {\sqrt{2}R}\sum^\infty_{w=1} \sum_{n\in\mathbb{Z}}\int_0^1 d\lambda\int_{-\infty}^\infty ds\, \eta(\hat{\tau})^{-4}e^{2\pi i\hat{\tau}(s^2-\frac{ w^2}{4}- w\lambda)}e^{2\pi i \hat{t} (\lambda+\frac{w}{2})}e^{\pi i \hat{\tau}(nR)^2} \\
& \qquad  \times \sum_{m\in\mathbb{Z}}\delta(\hat{t}-w\hat{\tau}-m)\,   \frac{1}{2}\Bigl[\prod_{n=1}^\infty (1+y\hat{q}^{n-\frac{1}{2}})(1+y^{-1}\hat{q}^{n-\frac{1}{2}})(1+\hat{q}^{n-\frac{1}{2}})^8 \\
& \qquad \qquad \quad +  
 \prod_{n=1}^\infty (1-y\hat{q}^{n-\frac{1}{2}})(1-y^{-1}\hat{q}^{n-\frac{1}{2}}) (1-\hat{q}^{n-\frac{1}{2}})^8   \\ & \qquad \qquad   \quad +16 \, \hat{q}^{\frac{5}{8}}(y^{\frac{1}{2}}+y^{-\frac{1}{2}})\prod_{n=1}^\infty (1+y\hat{q}^{n})(1+y^{-1}\hat{q}^{n})(1+\hat{q}^{n})^8 \Bigr] \ , 
    \stepcounter{equation}\tag{\theequation}
\end{align*} 
where we have written $y=e^{2\pi i m}$. The integral over $\lambda$ and the sum over $m$ can now be performed,
\begin{align} 
   &\int_0^1 d\lambda\, e^{-2\pi i w\hat{\tau} (\frac{w}{4}+\lambda)}e^{2\pi i \hat{t}(\lambda+\frac{w}{2})}\sum_{m\in\mathbb{Z}}\, \, \delta(\hat{t}-w\hat{\tau}-m)   \nonumber\\
    &=\hat{q}^{\frac{w^2}{4}}\int_0^1 d\lambda\, \sum_{m\in\mathbb{Z}}\, \, e^{2\pi i m(\lambda+\frac{w}{2})}\delta(\hat{t}-w\hat{\tau}-m)\\ 
   &= \hat{q}^{\frac{w^2}{4}} \sum_{m\in\mathbb{Z}}\,\delta_{m,0}\,e^{2\pi i m\frac{w}{2}}\delta(\hat{t}-w \hat{\tau}-m) 
   = \hat{q}^{\frac{w^2}{4}} \delta(\hat{t}-w \hat{\tau}) \nonumber  
 = \hat{q}^{\frac{w^2}{4}} \frac{1}{w}\, \delta\bigl(\tfrac{\hat{t}}{w}-\hat{\tau}\bigr) \ ,
\end{align}
where we have first replaced $\hat{t} = w\hat{\tau} + m$, and then performed the integral over $\lambda$, which localises the $m$-sum to $m=0$. Including the ghosts that remove two bosons and two fermions, we thus get for the physical spectrum 
\begin{align*}
\label{eq:closed_ampl1}
    \hat{Z}_{\rm phys}(\hat{t};\hat{\tau}) =& \sqrt{2}R\sum^\infty_{w=1} \sum_{n\in\mathbb{Z}}\int_{-\infty}^\infty ds\, \eta(\hat{\tau})^{-6}e^{2\pi i\hat{\tau}(s^2+\frac{w^2}{4})}e^{\pi i \hat{\tau}(nR)^2}\\
&\ \times \frac{1}{w}\delta(\tfrac{\hat{t}}{w}-\hat{\tau})\,  
\frac{1}{2}\left[\vartheta_3(\hat{\tau})^4+(-1)^{w+1} \vartheta_4(\hat{\tau})^4+\vartheta_2(\hat{\tau})^4\right] \ ,
    \stepcounter{equation}\tag{\theequation}
\end{align*}  
where we have rewritten the square bracket in terms of Jacobi theta functions. By the Jacobi abstruse identity, see eq.~(\ref{eq:abstruse}), the bracket now becomes
\be
\frac{1}{2}\left[\vartheta_3(\hat{\tau})^4+(-1)^{w+1} \vartheta_4(\hat{\tau})^4+\vartheta_2(\hat{\tau})^4\right] = \left\{ \begin{array}{ll} \vartheta_3(\hat{\tau})^4 &\quad  \hbox{if $w$ is odd} \\
\vartheta_2(\hat{\tau})^4 &\quad  \hbox{if $w$ is even} \ .
\end{array}
\right.
\ee
Upon doing the worldsheet integral over $\hat{\tau}$ we thus finally obtain 
\be
\label{eq:closed_final}
\int_0^{i\infty} d\hat{\tau} \hat{Z}_{\rm phys}(\hat{t};\hat{\tau}) = \sum_{w=1}^{\infty} \frac{1}{w} \hat{x}^{\frac{w}{4}}\, \hat{Z}_w(\hat{t}\, )\ , 
\ee
where $\hat{Z}_w(\hat{t}\, )$ equals 
\be
\label{eq:closed_final_w}
\hat{Z}_w(\hat{t}\, )= \sqrt{2}R \sum_{n\in\mathbb{Z}}\int_{-\infty}^\infty ds\, \, \eta\bigl(\tfrac{\hat{t}}{w}\bigr)^{-6}e^{2\pi i\frac{\hat{t}}{w}s^2}e^{\pi i \frac{\hat{t}}{w}(nR)^2}\vartheta_{\hat{*}(w)}\bigl(\tfrac{\hat{t}}{w}\bigr)^4 \ ,
\ee
and $\hat{*}(w) = 3$ for $w$ odd, and $\hat{*}(w) = 2$ for $w$ even. This result will be compared with the closed string cylinder diagram in the dual CFT, where it will account for the contributions from the single cycle sectors. (The contributions from multi-cycle sectors arise from including string field theoretic multi-string contributions.)

\subsection{The worldsheet open-string channel}

In order to translate this into the open string picture, we use the same transformation as above, see eq.~(\ref{openstring}). Then the amplitude of eq.~(\ref{eq:closed_ampl}) becomes
\begin{align*}
    {Z}_{\rm phys}({t};{\tau}) =& \sqrt{2}R\sum^\infty_{w=1} \sum_{n\in\mathbb{Z}}\int_{-\infty}^\infty ds\, \eta(-\tfrac{1}{\tau})^{-6}e^{ - 2\pi i\frac{1}{\tau}(s^2+\frac{w^2}{4})}e^{-\pi i \frac{1}{\tau}(nR)^2}\\
&\quad \times \frac{1}{w}\delta(-\tfrac{1}{wt}+\tfrac{1}{\tau})\,  
\frac{1}{2}\Bigl[\vartheta_3(-\tfrac{1}{\tau})^4+(-1)^{w+1} \vartheta_4(-\tfrac{1}{\tau})^4+\vartheta_2(-\tfrac{1}{\tau})^4\Bigr] \ . 
    \stepcounter{equation}\tag{\theequation}
\end{align*}
Next we rewrite the Dirac delta function for the open string variables as
\begin{equation}
    \frac{1}{w}\delta(-\tfrac{1}{wt}+\tfrac{1}{\tau}) =\frac{\tau^2}{w}\delta(\tau-wt)\ .
\end{equation}
Using the $S$-modular transformation formula for each term (and performing the integral over $s$) we then get 
\begin{equation}
    {Z}_{\rm phys}({t};{\tau}) = \sum^\infty_{w=1}\sum_{m\in\mathbb{Z}}\, \frac{\tau^2}{w} \,e^{-2\pi i\frac{1}{\tau}\frac{w^2}{4}}e^{\pi i {\tau}(\frac{m}{R})^2}\,\eta({\tau})^{-6}\,  \vartheta_{\star(w)}({\tau})^4\,\delta(\tau-wt) \ , 
\end{equation}
where $\star(w)=3$ for $w$ odd, and $\star(w)=4$ for $w$ even.
Finally, integrating over the modulus $\tau$,  leads to 
\begin{align}
\label{eq:open_final}
    \int_0^{i\infty}\frac{d\tau}{\tau^2} {Z}_{\rm phys}({t};{\tau}) &= \sum^\infty_{w=1}\sum_{m\in\mathbb{Z}}\, \frac{1}{w} \,\hat{x}^{\frac{w}{4}}e^{\pi i wt(\frac{m}{R})^2}\,\eta({wt})^{-6}\,  \vartheta_{\star(w)}({wt})^4  \\
    & = \sum_{w=1}^\infty \frac{1}{w} \hat{x}^{\frac{w}{4}} {Z}_{w}(t)  \ ,
\end{align}
where the measure is induced from (\ref{eq:closed_final}) upon the substitution $\hat{\tau} = -\frac{1}{\tau}$, see eq. (\ref{openstring}), and we have defined (recall that $\hat{x} = e^{2\pi i \hat{t}} = e^{-\frac{2\pi i}{t}}$)
\begin{align}\label{3.43}
    {Z}_{w}(t)  &= \sum_{m\in\mathbb{Z}} e^{\pi i {wt}(\frac{m}{R})^2}\,\eta({wt})^{-6}\,  \vartheta_{\star(w)}({wt})^4\ .
\end{align}
The minus signs that appear in the open string for even $w$ (for which $\star(w)=4$) are due to the fermions; in fact, as we shall see, they are required to reproduce the open string channel in the dual CFT, see the discussion below eq.~(\ref{eq:openstring}).

\subsubsection*{Open strings between branes with different spin-structure}
We can similarly calculate the relative overlap between two boundary states of opposite spin structure
\begin{equation}
    \hat{Z}^-(\hat{t};\hat{\tau}) =\,_{\text{S}}\langle\!\langle W',\Lambda',S',\theta',\mp\,\| \hat{q}^{\frac{1}{2}(L_0+\bar{L}_0-\frac{c}{12})}\hat{x}^{\frac{1}{2}(J_0^3+\bar{J}_0^3)}\| W,\Lambda,S,\theta,\pm\,\rangle\!\rangle_\mathrm{S} \ ,
\end{equation}
where the superscript $-$ is used to distinguish this case from the previous calculation. For the case that $W'=W$, $\Lambda'=\Lambda$, $S'=S$ and $\theta'=\theta$ this then leads in the closed channel to 
\begin{align}
\label{eq:closed_ampl_rel1}
    \hat{Z}^-_{\rm phys}(\hat{t};\hat{\tau}) =& \sqrt{2}R\sum^\infty_{w=1} \sum_{n\in\mathbb{Z}}\int_{-\infty}^\infty ds\, \eta(\hat{\tau})^{-6}e^{2\pi i\hat{\tau}(s^2+\frac{w^2}{4})}e^{\pi i \hat{\tau}(nR)^2} \nonumber \\
&\quad \times \frac{1}{w}\delta(\tfrac{\hat{t}}{w}-\hat{\tau})\,  
\frac{1}{2}\Bigl[\vartheta_3(\hat{\tau})^4+(-1)^{w+1} \vartheta_4(\hat{\tau})^4-\vartheta_2(\hat{\tau})^4\Bigr] \ ,   
\end{align}
i.e.\ the only difference from eq.~(\ref{eq:closed_ampl}) is the sign in front of the $\vartheta_2$-term, see eq.~(\ref{eq:GSOstate}). Now the abstruse identity leads to
\begin{align}
\label{eq:closed_ampl_rel2}
    \hat{Z}^-_{\text{phys}}(\hat{t};\hat{\tau}) =& \sqrt{2}R\sum^\infty_{w=1}\sum_{n\in\mathbb{Z}}\int_{-\infty}^\infty ds\,  \eta(\hat{\tau})^{-6}e^{2\pi i\hat{\tau}({s^2}+\frac{ w^2}{4})}e^{\pi i \hat{\tau}(nR)^2} \frac{1}{w}\delta(\hat{t}-w\hat{\tau})\vartheta_{\hat{*}'(w)}(\hat{\tau})^4 \ , 
    \end{align}
where $\hat{*}'(w) = 4$ for $w$ odd, and $\hat{*}'(w) = 1$ for $w$ even. (Note that this effectively removes the even $w$ sector since $\vartheta_1=0$.) In terms of the open string variables, see eq.~(\ref{eq:closed_ampl_rel1}), this then becomes 
\begin{align}
    \int_0^{i\infty}\frac{d\tau}{\tau^2}\,  {Z}^-_{\rm phys}({t};{\tau}) & = \sum^\infty_{w=1}\sum_{m\in\mathbb{Z}}\, \frac{1}{w} \,\hat{x}^{\frac{w}{4}}e^{\pi i {wt}(\frac{m}{R})^2}\,\eta({wt})^{-6}\,  \vartheta_{\star'(w)}({wt})^4 \\
    & = \sum_{w=1}^\infty \frac{1}{w} \hat{x}^{\frac{w}{4}} {Z}^-_{w}(t)  \ ,
\end{align}
where $\star'(w) = 2$ for $w$ odd, and $\star'(w) = 1$ for $w$ even, and  
\begin{align}\label{3.48}
    {Z}^-_{w}(t)  &= \sum_{m\in\mathbb{Z}}e^{\pi i {wt}(\frac{m}{R})^2}\,\eta({wt})^{-6}\,  \vartheta_{\star'(w)}({wt})^4\ .
\end{align}

\section{The boundary states of the  dual CFT}
\label{Sec:Orbifold}

It was shown in \cite{Gaberdiel:2024dva} that the dual CFT to the above worldsheet theory is the symmetric product orbifold of $8$ free fermions and $2$ free bosons. As in \cite{Gaberdiel:2021kkp} we should therefore expect to be able to identify the above D-branes with suitable boundary states of this symmetric orbifold CFT.\footnote{For other recent discussions of boundary states and defects in symmetric orbifold theories, see also \cite{Belin:2021nck,Harris:2025wak}.} Many of the results from \cite{Gaberdiel:2021kkp} carry over, and we can therefore be relatively brief. 

In particular, as in \cite{Gaberdiel:2021kkp}, the above worldsheet boundary states correspond to the maximally fractional boundary states of the symmetric orbifold. These are characterised by the property that they satisfy `factorised' gluing conditions, i.e.\ that the boundary conditions only identify left- and right-moving fields from the same copy. In particular, these maximally symmetric boundary states therefore 
descend from boundary states of the seed theory, and we shall therefore discuss these first.

\subsection{The boundary states of the seed theory}

As mentioned above, the seed theory of the dual symmetric orbifold consists of $8$ free fermions and two free bosons.\footnote{It is therefore equal to $({\cal S}_0)^2$, where ${\cal S}_0$ is the simplest example of a large ${\cal N}=4$ model, consisting of one free boson and four free fermions, see e.g.\ \cite{Gukov:2004ym}.} One of the bosons corresponds directly to the ${\rm S}^1$ boson that is part of the ${\rm AdS}_3\times {\rm S}^3 \times {\rm S}^3 \times {\rm S}^1$ background, while the other boson is associated to the radial direction of ${\rm AdS}_3$. While its precise nature could not be identified in \cite{Gaberdiel:2024dva}, it follows from the subsequent analysis of \cite{Eberhardt:2025sbi} that it is in fact a simple free boson (with continuous spectrum). We shall denote the modes of the latter boson by $\alpha^0_n$ and $\bar{\alpha}^0_n$, whereas the modes of the compact boson will be denoted by $\alpha^1_n$ and $\bar{\alpha}^1_n$. Mirroring what we did in the bulk, we shall impose the following gluing conditions on these seed theory degrees of freedom 
\begin{subequations}
\label{eq:IshibashiOrbifold}
\begin{align}
    (\alpha_n^0-\bar{\alpha}_{-n}^0)|p, (p_L,p_R),\varepsilon\rangle\!\rangle_{\mathrm{S}} &=0\ , \label{eq:R0} \\
    (\alpha_n^1+\bar{\alpha}_{-n}^1)|p, (p_L,p_R),\varepsilon\rangle\!\rangle_{\mathrm{S}} &=0\ , \label{eq:R1} \\
(\chi_r^{j}+i\varepsilon\delta^j\bar{\chi}_{-r}^{j})|p, (p_L,p_R), \varepsilon \rangle\!\rangle_{\mathrm{S}} &=0\ ,\ \label{gluingfermionsseed}
\end{align}
\label{eq:gluingseed}
\end{subequations}

\vspace{-14pt}
\noindent where $j=1,\ldots,8$ labels the different fermions, and $p$ is the momentum eigenvalue of the non-compact boson, while $(p_L,p_R)$ denote the left- and right-moving momenta of the ${\rm S}^1$ boson. Furthermore, $\delta^j = +1$ for $j=1,2,3,4$, say, and $\delta^j=-1$ for $j=5,6,7,8$. This reflects that, after removing the light-cone directions, we have $4$ $+$ signs and $4$ $-$ signs for the fermionic gluing conditions on the worldsheet. This assignment is also natural from the perspective of the $({\cal S}_0)^2$ theory, since each ${\cal S}_0$ factor contains one boson and four free fermions, and we impose opposite boundary conditions (Neumann vs.\ Dirichlet) on the two ${\cal S}_0$ factors. (The fermionic gluing conditions follow from the bosonic gluing conditions by requiring large ${\cal N}=4$ superconformal symmetry.) Indeed, in order to match with what we did on the worldsheet we  have imposed a Neumann boundary condition on the compact boson ${\rm S}^1$, while for the non-compact radial direction (corresponding to $J^3$) we have imposed a Dirichlet boundary condition. 

The the compact ${\rm S}^1$ boson, the Ishibashi states only arise for $p_L=-p_R$. (This simply follows from eq.~(\ref{eq:R1}) with $n=0$.) For a circle of radius $R$, these eigenvalues are of the form $p_L = - p_R = n R$. Finally, given that we shall ultimately be interested in the symmetric orbifold of this seed theory, we will only need to consider the NS sector of the fermions.\footnote{The Ramond sector of the symmetric orbifold has ground state conformal dimension proportional to $N$, and hence decouples in the large $N$ limit.} 
Out of these Ishibashi states we can then form consistent boundary states, and they can be factorised 
into the product of two bosons and 8 fermions mirroring what we did in eq.~(\ref{eq:fullBS}),
\be
|\!| x,\theta,\pm\rangle\!\rangle  = |\!|x\rangle\!\rangle \otimes  |\!|\theta\rangle\!\rangle \otimes |\!|\pm\rangle\!\rangle \ , 
\ee
where $|\!|x\rangle\!\rangle$ denotes the boundary state of the non-compact boson, see eq.~(\ref{eq:BS_R0}), $ |\!|\theta\rangle\!\rangle$ the boundary state of the compact boson, see eq.~(\ref{Dirbrane}), while the fermionic boundary state $|\!|\varepsilon\rangle\!\rangle = | {\rm D}/{\rm N},\varepsilon \rangle\!\rangle_{\rm NS} $ with $\varepsilon=\pm$ is described in Appendix~\ref{appFermions}. (Recall that $4$ of the fermions satisfy Dirichlet, and $4$ Neumann boundary conditions.)
 
The overlaps of these boundary states are calculated in Appendix~\ref{app:WS}, and the corresponding open string spectra are simply 
\begin{align}
& \langle\!\langle y,\theta_2,\pm\|\, e^{\pi i \hat{t}(L_0+\bar{L}_0-\frac{c}{12})}\,\| x,\theta_1,\pm\rangle\!\rangle \\
& \qquad = R\sum_{n\in\mathbb{Z}}\int_{-\infty}^\infty d{p}\, e^{\pi in(\theta_1-\theta_2)}\,e^{2\pi i {p}(x-y)} \, \chi_{p}({\hat{t}}\,)\chi_{nR} ({\hat{t}}\,) \frac{\vartheta_3(\hat{t}\,)^4}{\eta(\hat{t}\,)^4}\nonumber\\ 
    & \qquad = \sum_{m\in\mathbb{Z}} \chi_{(x-y)}(t)\, \chi_{\frac{1}{2R}(2m+\theta_1-\theta_2)}(t) \, \frac{\vartheta_3(t)^4}{\eta(t)^4} \ , \label{seed_open}
\end{align}
where $\chi_p(\hat{t}\,)$ is defined in eq.~(\ref{A.10}), and we have used the usual translation to the open string coordinate,
\be
t = - \frac{1}{\hat{t}} \ . 
\ee
The resulting spectrum in (\ref{seed_open}) describes an open string stretching between positions $x$ and $y$ in the non-compact direction. Furthermore, along the compact ${\rm S}^1$, the relative Wilson line shifts the momentum modes (that are parametrised by $m$) by $\frac{\theta_1-\theta_2}{2R}$.

\subsection{Boundary states in the symmetric orbifold}

The boundary states that are relevant for the above ${\rm AdS}_3$ D-branes are the `maximally fractional' ones for which one imposes the above gluing conditions (\ref{eq:IshibashiOrbifold}) separately for each copy.\footnote{As a consequence, the gluing conditions are then also the same in the twisted sectors.} Then the  corresponding Ishibashi states exist in every twisted sector, and the boundary states take the form, see \cite{Gaberdiel:2021kkp,Belin:2021nck}
\begin{equation}
\label{maxfracbrane}
\| x, \theta , \rho, \varepsilon \rangle\!\rangle =\sum_{[\sigma]}  
\Bigl( \tfrac{|[\sigma]|}{N!} \Bigr)^{\frac{1}{2}}\, \chi_\rho([\sigma])\, 
 R^{\frac{r}{2}}\prod_{j=1}^{r} \sum_{n_j\in\mathbb{Z}}\int_{\mathbb{R}} d{p}_{j} \, e^{2\pi i {p}_{j} x}e^{\pi i{n}_j\theta}
\big|  {\mathbf{p}}, \mathbf{n}R, \varepsilon \big\rangle \!\big\rangle_{[\sigma]} \ .
\end{equation}
Here the Ishibashi state $|  {\mathbf{p}}, \mathbf{n}R, \varepsilon \rangle \!\rangle_{[\sigma]} $ is defined in the $[\sigma]$-twisted sector, where \linebreak $\mathbf{p} = (p_{1},\ldots,p_{r})$ and $\mathbf{n} = (n_{1},\ldots, n_{r})$, with $r$ the number of cycles in the conjugacy class $[\sigma]$. Furthermore,  $\rho$ is a representation of $S_N$, and $|[\sigma]|$ denotes the number of elements in the conjugacy class $[\sigma]$. 

As mentioned before, the fermions are all taken to be in the NS sector, and the overlap of the (twisted) Ishibashi states with the same spin structure is 
\begin{equation}
{}_{[\sigma]}\big\langle\!\big\langle {\mathbf{p}},\mathbf{n}R,\pm\big|\, e^{\pi i \hat{t}(L_0+\bar{L}_0-\frac{Nc}{12})}\,\big| {\mathbf{p}},\mathbf{n}R,\pm\big\rangle\!\big\rangle_{[\sigma]} =
\prod_{j=1}^{r} \, \chi_{{p}_{j}} \left(\tfrac{\hat{t}}{l_j}\right)\, \chi_{n_jR} \left(\tfrac{\hat{t}}{l_j}\right) \,
 \frac{\vartheta_{*(l_j)}(\frac{\hat{t}}{l_j})^4}{\eta(\frac{\hat{t}}{l_j})^4} \ , 
 \label{eq:BasicOverlap}
\end{equation}
where the lengths of the $r$ cycles in $[\sigma]$ are $l_1,\ldots,l_r$, corresponding to the partition $N = \sum_{j=1}^r l_j$, and $*(l_j) = 3$ if $l_j$ is odd, and $*(l_j)=2$ if $l_j$ is even. The overlap between two such boundary states is then 
\begin{align}
&  \langle\!\langle y,\theta_2,\rho_2, \pm\|\, e^{\pi i \hat{t}(L_0+\bar{L}_0-\frac{c}{12})}\,
\| x, \theta_1 , \rho_1, \pm \rangle\!\rangle  \nonumber \\
&\quad  = \sum_{\sigma\in S_N}\frac{R^{\,r}}{ N!}\bar{\chi}_{\rho_1}([\sigma]){\chi}_{\rho_2}([\sigma]) \nonumber \\
&\quad \qquad \qquad \times 
\prod_{j=1}^{r} \,\sum_{n_j\in\mathbb{Z}}\int_{\mathbb{R}} d{p}_{j} \, e^{2\pi i {p}_{j}\Delta_{x,y}}e^{\pi i{n}_j\Delta_{\theta}}\,
\chi_{{p}_{j}} \bigl(\tfrac{\hat{t}}{l_j}\bigr)\, \chi_{n_jR} \bigl(\tfrac{\hat{t}}{l_j}\bigr) \, 
 \frac{\vartheta_{\hat{*}(l_j)}(\frac{\hat{t}}{l_j})^4}{\eta(\frac{\hat{t}}{l_j})^4} \label{eq:overfirstorb}
\\
&\quad =\frac{1}{N!}\sum_{\sigma\in S_N}\bar{\chi}_{\rho_1}([\sigma]){\chi}_{\rho_2}([\sigma]) \prod_{j=1}^{r} \,\,\sum_{m_j\in\mathbb{Z}} \chi_{\Delta_{x,y}}(l_j t) \, 
\chi_{\frac{1}{2R}(2m_j+\Delta_{\theta})} (l_jt) \,   \frac{\vartheta_{\star(l_j)}(l_jt)^4}{\eta(l_jt)^4} \nonumber \\
&\quad  = \,\sum_{m\in\mathbb{Z}} \left[\frac{1}{N!}\sum_{\sigma\in S_N}\bar{\chi}_{\rho_1}([\sigma]){\chi}_{\rho_2}([\sigma]) \mathrm{Tr}_{\hat{\mathcal{H}}_m^{\otimes N}} \left(\sigma e^{2\pi i t (L_0 - \frac{Nc}{24})}\right) \right] \ , \label{eq:openstring}
\end{align} 
where $\hat{*}(l_j) =\star(l_j)= 3$ if $l_j$ is odd, and $\hat{*}(l_j)=2$ and $\star(l_j)=4$ if $l_j$ is even. Furthermore, $\Delta_{x,y} = x-y$ and $\Delta_{\theta} = \theta_1-\theta_2$, and $\hat{\mathcal{H}}$ denotes the open string 
spectrum between two D-branes at relative position $x-y$ in the non-compact direction, and with relative Wilson line $\theta_1-\theta_2$ in the compact direction. (As before, $m$ labels the momentum modes along the compact circle, see eq.~(\ref{A.6}).) Finally, in going to the last line we have used that a cyclic permutation of even length is odd, and hence introduces a minus sign on the fermionic excitations --- this is the reason why we have $\star(l_j)=4$ for $l_j$ even. 

The sum over the permutations $\sigma$ now projects the open string spectrum onto those representations that appear in the tensor product of $\rho_1 \otimes \rho_2^*$; in particular, for $\rho_1=\rho_2={\rm id}$, the trivial representation, the open string spectrum is projected onto the singlet states. 

The situation where we consider two D-branes with opposite spin structure proceeds analogously: we only need to change the assignment of $\hat{*}(l_j)$  and $\star(l_j)$: if $l_j$ is odd, we now get $\hat{*}(l_j) = 4$ (instead of $\hat{*}(l_j) = 3$, reflecting the relative spin structure sign) and hence $\star(l_j)=2$; whereas for $l_j$ even we find $\hat{*}(l_j)=1$ (instead of $\hat{*}(l_j)=2$), and hence also $\star(l_j)=1$.  In terms of the open string perspective this simply means that now the relative open string is in the Ramond sector.

\subsection{Identification with worldsheet branes and tests}

It remains to identify these symmetric orbifold boundary states with what we found in the worldsheet theory. Our basic claim is that the worldsheet boundary states translate directly to the symmetric orbifold constructions with $\rho=\mathrm{id}$. In the following we shall demonstrate that, with this identification, the closed string calculations match; in the following section we shall explain a similar matching for the open string.

We start by rewriting the closed string version of the symmetric orbifold overlap, see eq.~(\ref{eq:overfirstorb}), as  
\begin{align}
\label{eq:overfirstorb_closed}
    \hat{Z}^{S_N}_{x,\theta_1|y,\theta_2}(\hat{t}\,) & =  \langle\!\langle y,\theta_2,\mathrm{id}, \pm\|\, e^{\pi i \hat{t}(L_0+\bar{L}_0-\frac{c}{12})}\, \| x, \theta_1 , \mathrm{id}, \pm \rangle\!\rangle \\
& =
\frac{1}{N!}\sum_{\sigma\in S_N} \prod_{j=1}^{r} \hat{Z}_{x,\theta_1|y,\theta_2;\,\hat{s}(l_j)}(\tfrac{\hat{t}}{l_j}) \, \ ,
\end{align}
with $\hat{s}(l_j) = {\mathrm{NS}}$ for odd $l_j$ and $\hat{s}(l_j) = {\mathrm{R}}$ for even $l_j$. From now on, we will suppress the labels $x$, $y$, $\theta_1$ and $\theta_2$, and in fact set them to zero for simplicity --- the calculation for the general case works similarly. For each $j$, we then have 
\begin{align}
\label{eq:closedchannelplus}
 &\hat{Z}_{\hat{s}(l_j)}\left(\tfrac{\hat{t}}{l_j}\right)= R\sum_{n_j\in\mathbb{Z}}\int_{-\infty}^\infty dp_j\,
 \chi_{{p}_{j}} \left(\tfrac{\hat{t}}{l_j}\right)\, \chi_{n_jR} \left(\tfrac{\hat{t}}{l_j}\right) \,
 \frac{\vartheta_{\hat{*}(l_j)}(\frac{\hat{t}}{l_j})^4}{\eta(\frac{\hat{t}}{l_j})^4} \ .   
\end{align}
Next we recall that eq.~(\ref{eq:overfirstorb_closed}) gives the full symmetric orbifold description, whereas the worldsheet string theory only captures the single-particle contributions. In order to relate the two descriptions we pass to the grand canonical ensemble, following \cite{Eberhardt:2020bgq}, i.e.\ we sum over all values of $N$ with a chemical potential $\hat{p}=e^{2\pi i \sigma}$, 
\begin{align}
\hat{\mathfrak{Z}}(\hat{p},\hat{t}\,)&=\sum_{N=1}^\infty \hat{p}^N  \hat{Z}^{S_N}(\hat{t}\,)\ .
\end{align}
In order to extract the contribution from a single worldsheet we then take the logarithm of this expression, leading to 
\begin{align}
\log\left(\hat{\mathfrak{Z}}(\hat{p},\hat{t}\,)\right)&
= \sum_{\substack{w=1\\ \text{$w$ odd}}}^\infty \frac{\hat{p}^w}{w} 
\hat{Z}_{\mathrm{NS}}\left(\tfrac{\hat{t}}{w}\right) 
+ \sum_{\substack{w=1\\ \text{$w$ even}}}^\infty 	
\frac{\hat{p}^w}{w} \hat{Z}_{\mathrm{R}}\left(\tfrac{\hat{t}}{w}\right)  \\
& = \sum_{w=1}^{\infty} \frac{\hat{p}^w}{w} R\sum_{n\in\mathbb{Z}}\int_{-\infty}^\infty dp\,
	\chi_{p} \left(\tfrac{\hat{t}}{w}\right)\, \chi_{nR} \left(\tfrac{\hat{t}}{w}\right)
	\frac{\vartheta_{\hat{*}(w)}(\frac{\hat{t}}{w})^4}{\eta(\frac{\hat{t}}{w})^4} \ , 
\label{eq:orbi_closed+}
\end{align}
where again $\hat{*}(w)=3$ for odd $w$ and $\hat{*}(w)=2$ for even $w$. Each term in the sum matches now with the result from the worldsheet theory in eq.~(\ref{eq:closed_final_w}), except for the factor of $\hat{x}^{\frac{w}{4}}$ in eq.~(\ref{eq:closed_final}); this overall factor is corrected by the contribution of disconnected discs (which are difficult to compute), and hence cannot be easily compared. 

\subsubsection*{The open string perspective}

We can also translate these closed string expressions into the open string channel by applying the $S$-modular transformation. In particular, this will turn the $\hat{*}(w)=3$ NS-characters (for odd $w$) into NS-characters, while the $\hat{*}(w)=2$ R-like characters (for even $w$) become NS-characters with the insertion of  $(-1)^F$. From the perspective of the symmetric orbifold this is natural since fermionic states pick up a sign under an even $w$ cycle permutation. In fact, the same contributions appear on the worldsheet, see eq.~(\ref{3.43}), and hence there is again a precise match, as expected.

\subsubsection*{The case of opposite spin structures}

Finally, the analysis for two D-branes with opposite spin structure is essentially the same: the only difference is that now the open string in the symmetric orbifold will be in the R sector (for odd $w$), resp.\ in the R sector with $(-1)^F$ for even $w$. This matches precisely with what we found in the worldsheet description, see eq.~(\ref{3.48}).

\section{${\rm AdS}_2$ branes}\label{sec:AdS2}

In this section we comment briefly on the construction of the boundary states corresponding to $\mathrm{AdS}_2$ branes in ${\rm AdS}_3 \times {\rm S}^3 \times {\rm S}^3 \times {\rm S}^1$. As argued in \cite{Harris:2025wak}, these branes are dual to interfaces involving reflective and transmissive boundaries in the symmetric orbifold. As we will see, we can also reproduce these results in the NS-R description. 

Let us begin by discussing the gluing conditions for the $\mathrm{AdS}_2$ branes: they take the form 
\begin{subequations}
\label{eq:IshibashiAdS2}
\begin{align}
    (J_n^a+\bar{J}_{-n}^a)|0,\lambda,s,\varepsilon\rangle\!\rangle_{\rm AdS}  &=0\ , \label{J3condAdS2}\\
    (K_n^{\, a (\pm)}+\bar{K}_{-n}^{\, a (\pm)})|0,\lambda,s,\varepsilon\rangle\!\rangle_{\rm AdS}  &=0\ ,\\
    (\psi_r^a+i\varepsilon \,  \bar{\psi}_{-r}^a)|0,\lambda,s,\varepsilon\rangle\!\rangle_{\rm AdS}  &=0\ , \label{eq:gc_fermAdS2}\\
    (\chi_r^{\, a (\pm)}+i\varepsilon \,  \bar{\chi}_{-r}^{\, a (\pm)})|0,\lambda,s,\varepsilon\rangle\!\rangle_{\rm AdS}  &=0\ ,\\
    (\alpha_n^1+\bar{\alpha}_{-n}^1)|0,\lambda,s,\varepsilon\rangle\!\rangle_{\rm AdS}  &=0\ ,  \label{eq:gc_circlebAdS2} \\
    (\psi_r^1+i\varepsilon \,  \bar{\psi}_{-r}^1)|0,\lambda,s,\varepsilon\rangle\!\rangle_{\rm AdS}  &=0\ . 
    \label{eq:gc_circleAdS2}\
\end{align}
\end{subequations}
As was already noticed in \cite{Gaberdiel:2021kkp}, the condition on $J^3_0$, implies $m = -\bar{m}$, where $m = \lambda + \mathbb{Z}$ is the eigenvalue of $J^3_0$. This restricts $\lambda$ to the values $\{0, \frac{1}{2}\}$, and excludes any non-trivial spectral flow, i.e.\ the only non-trivial Ishibashi states come from the $w = 0$ sector. As before, see footnote~\ref{foot5}, we have mirrored the gluing conditions for ${\rm AdS}_3$ by the two ${\rm S}^3$'s, and as a consequence now all ten fermions have a $+i\varepsilon$ gluing condition. Thus all the $8$ fermions of the dual CFT satisfy the same (Neumann-like) gluing condition. Furthermore, the non-compact boson (corresponding to the $J^3$ direction) now also satisfies a Neumann condition; as a consequence, the normalisation of the corresponding boundary state is a bit delicate --- in the description of \cite{Harris:2025wak} this is reflected in the fact that the normalisation factor ${\cal N}$ below their eq.~(4.5) equals $t$. We also note that the gluing conditions for the fermions are compatible with the GSO projection for Type IIB (since all the ten fermions satisfy the same $+i\varepsilon$ condition).

Following the proposal of \cite{Gaberdiel:2021kkp}, the ansatz for the boundary states corresponding to the ${\rm AdS}_2$ branes is 
\begin{align}\label{5.2}
\|S,\theta,\pm\rangle\!\rangle_\mathrm{A} = 2^{-\frac{1}{4}}\sum_{\lambda=0,\frac{1}{2}}\int_{\mathbb{R}} ds\, e^{2\pi i sS}e^{2\pi i(\lambda-\frac{1}{2})\delta^L_A}\, \, |0,\lambda,s\rangle\!\rangle_{\rm AdS} \,  \otimes \|\theta\rangle\!\rangle \,  \otimes \| {\rm F}, \pm \rangle\!\rangle_A \ ,
\end{align}
where $S\in\mathbb{R}$, and the continuous parameter $s$ takes value in the real line, $s\in\mathbb{R}$. These states describe half-branes, indicated by the label $A\in \{L,R\}$.  As before, see the discussion below eq.~(\ref{eq:GSOstateNS}), we have combined the different spin structures for the fermions via 
\be
 \| {\rm F}, \pm \rangle\!\rangle_A = \| {\rm NS} \rangle\!\rangle \pm e^{\pi i \delta^L_A}\| {\rm R} \rangle\!\rangle \ ,
\ee
where 
\begin{align}
	\| {\rm NS} \rangle\!\rangle &= \tfrac{1}{2}\big[| {\rm NS}, +\rangle\!\rangle + | {\rm NS}, -\rangle\!\rangle\big]  \ , \\
	\| {\rm R} \rangle\!\rangle &= \tfrac{1}{2}\big[| {\rm R},+\rangle\!\rangle + | {\rm R},-\rangle\!\rangle\big] \ .
\end{align}
In the Ramond sector, the phase factor of $e^{\pi i \delta^L_A}$ compensates for the shift of the $J^3_0$ eigenvalues by $\pm \frac{1}{2}$ arising from the fermionic zero modes; this effectively exchanges the roles of the sectors with $\lambda = 0$ and $\lambda= \frac{1}{2}$. 
Finally,  $\|\theta\rangle\!\rangle$ denotes the Neumann boundary state with Wilson line $\theta$ along the ${\rm S}^1$ as in Section~\ref{sec:cylinder}. The overlap of two such boundary states is then given by 
\begin{align}
& _{B}\langle\!\langle S_2,\theta, \pm |\!| \hat{q}^{\frac{1}{2}(L_0 + \bar{L}_0 - \frac{c}{12})}\, \hat{x}^{\frac{1}{2}(J^3_0 - \bar{J}^3_0)} |\!| S_1,\theta,\pm \rangle\!\rangle_A \nonumber \\
& \qquad = \frac{R}{\sqrt{2}} \sum_{\lambda=0,\frac{1}{2}}\sum_{n\in\mathbb{Z}} \int_{\mathbb{R}} ds\, 
e^{2\pi i s (S_1 - S_2)} e^{2\pi i (\lambda-\frac{1}{2}) (\delta_A^L - \delta_B^L)}\,  {\rm ch}[(\mathscr{F}_{\lambda,s})](\hat{t};\hat{\tau}) \,\chi_{nR}(\hat{\tau})\nonumber \\
& \hspace{6em} \times\, \left[ \frac{\vartheta_3(\hat{t},\hat{\tau})^5}{\eta(\hat{\tau})^{5}} + \frac{\vartheta_4(\hat{t},\hat{\tau})^5}{\eta(\hat{\tau})^{5}}+ e^{\pi i (\delta^L_A-\delta^L_B)}\frac{\vartheta_2(\hat{t},\hat{\tau})^5}{\eta(\hat{\tau})^{5}}\right] \\
& \qquad = \frac{R}{\sqrt{2}} \sum_{n,m\in\mathbb{Z}} \int_{\mathbb{R}} ds\, 
e^{2\pi is  (s\hat{\tau}+ S_1 - S_2)} \frac{e^{\pi i m} - e^{\pi i \delta_A^B}}{\eta(\hat{\tau})^8} \, \delta(\hat{t}-m)\,\chi_{nR}(\hat{\tau})  \nonumber\\ 
& \hspace{6em} \times\, \left[\vartheta_3(\hat{\tau})^5+\vartheta_4(\hat{\tau})^5+\vartheta_2(\hat{\tau})^5\right] \,  \ , \nonumber 
\end{align} 
where we have used eq.~(\ref{eq:characterSL}) with $w=0$ in the final step. This mirrors then precisely eq.~(3.17) in \cite{Harris:2025wak}. 

As suggested in \cite{Harris:2025wak}, it is natural to sum the two halves ($A=L$ and $A=R$), and the open string spectrum of this sum matches then with the partition function of the interface changing operators proposed in \cite{Harris:2025wak}. 

\section{Conclusions}\label{Sec:conclusions}

In this paper we have repeated the analysis of the spherical D-branes of \cite{Gaberdiel:2021kkp} for the case of ${\rm AdS}_3\times {\rm S}^3 \times {\rm S}^3 \times {\rm S}^1$ that was recently shown to be dual to the symmetric orbifold of eight free fermions and two bosons \cite{Gaberdiel:2024dva,Eberhardt:2025sbi}. The main novelty relative to the analysis of \cite{Gaberdiel:2021kkp} is that the worldsheet theory for  ${\rm AdS}_3\times {\rm S}^3 \times {\rm S}^3 \times {\rm S}^1$ can be described in the NS-R formalism, whereas the original analysis of \cite{Gaberdiel:2021kkp} for the case of ${\rm AdS}_3\times {\rm S}^3 \times \mathbb{T}^4$  was done in the hybrid description (since the NS-R description of that background is somewhat problematic). As a consequence, the flavour of the analysis is somewhat different since the worldsheet theory needs to be GSO projected. The fact that everything works out as expected is therefore another consistency check of the proposal of  \cite{Gaberdiel:2024dva,Eberhardt:2025sbi}.

We have mainly concentrated on the spherical D-branes for which the dual CFT description involves maximally fractional boundary states as in \cite{Gaberdiel:2021kkp}. We have also sketched the analysis for the ${\rm AdS}_2$ branes, see Section~\ref{sec:AdS2}, and our results agree again with those discussed in \cite{Gaberdiel:2021kkp}, and are compatible with the dual CFT interpretation proposed in \cite{Harris:2025wak}. It would be interesting to understand whether these results throw any light on the worldsheet description of quantum information theoretic aspects such as, e.g.\  entanglement entropy \cite{Ryu:2006bv}.

\section*{Acknowledgments}
We thank Vit Sriprachyakul for useful discussions. 
The work of MRG is supported by a personal grant from the Swiss National Science Foundation, and the work of his group at ETH is also supported in part by the Simons Foundation grant 994306  (Simons Collaboration on Confinement and QCD Strings), as well as the NCCR SwissMAP that is also funded by the Swiss National Science Foundation.

\appendix
\label{appendix}

\section{Boundary states}
\label{app:WS}

In this Appendix we collect the relevant formulae for the boundary states of the various factors. 

\subsection{Boundary states on the circle}
\label{appCircle}
The chiral character for a compact boson on a circle of radius $R$ is given by
\begin{equation}
\chi(\tau)=\frac{\Theta_\Lambda(\tau)}{\eta(\tau)} \ ,
\end{equation}
where 
\begin{equation}
\Theta_\Lambda(\tau) = \sum_{p_L\in\Lambda}q^{\frac{1}{2}p_L^2} \ , \quad q=e^{2\pi i \tau} \ ,
\end{equation}
and $\Lambda$ denotes the lattice of momentum and winding eigenvalues with 
\be
p_L = \left(\frac{m}{2R}+nR\right) \ , \qquad p_R = \left(\frac{m}{2R}-nR\right) \ , 
\ee
and $m,n\in\mathbb{Z}$. (For the corresponding right-movers the sum runs then over $p_R\in\Lambda$.) We shall be considering Neumann branes, i.e.\ the gluing condition (\ref{eq:gc_circleb}), resp.\ (\ref{eq:R1}). As a consequence, we need to impose the condition $p_L= - p_R$, i.e.\ only the terms with $m=0$ contribute. There is a one-parameter family of such D-branes, parametrised by the Wilson line $\theta$ on the circle, and the explicit form of the boundary state $|\!|\theta \rangle\!\rangle$ is 
\begin{equation}
\label{Dirbrane}
|\!|\theta \rangle\!\rangle =  \sqrt{R}\,
\sum_{n\in\mathds{Z}} e^{\pi i n \theta } \, 
\vert \, (nR,- nR)\, \rangle\!\rangle_{\rm N}\ ,
\end{equation}
where the Ishibashi state on the right-hand-side satisfies the gluing condition (\ref{eq:gc_circleb}) for all $n$. (This fixes it uniquely up to normalisation.) The closed string overlap between two such boundary states is then
\begin{align}
\langle\!\langle \theta_1 |\!| e^{{\pi i \hat{\tau}(L_0+\bar{L}_0-\frac{1}{12})}} |\!|\theta_2 \rangle\!\rangle & = R\, \sum_{n\in\mathbb{Z}} e^{\pi i n (\theta_2-\theta_1)} e^{\pi i \hat{\tau} n^2 R^2}\, \frac{1}{\eta(\hat{\tau} )}\ , \\
& = \sum_{m\in\mathbb{Z}} \, e^{\pi i \tau \frac{(\theta_2 - \theta_1+ 2m)^2}{4 R^2}}  \frac{1}{\eta(\tau)}\ , \label{A.6}
\end{align}
where $\tau = -\frac{1}{\hat{\tau}}$ is the open string modular parameter, and our conventions for the $\eta$ function are spelled out in Appendix~\ref{app:theta}. For the $S$-modular transformation we employed the Poisson resummation formula, i.e.
\be
 \sum_{n\in\mathbb{Z}} e^{-\pi a n^2 +bn} = \frac{1}{\sqrt{a}}\sum_{m\in\mathbb{Z}} e^{-\frac{\pi}{a}(m+\frac{b}{2\pi i})^2} \ .
\ee
We note that the open string spectrum in eq.~(\ref{A.6}) consists of open strings with momentum $\frac{m}{R}$, where $m \in \frac{1}{2} (\theta_2-\theta_1) + \mathbb{Z}$. This is as expected for the open string between two Neumann branes with Wilson lines $\theta_1$ and $\theta_2$.

\subsection{Boundary states for a non-compact boson}
\label{appR0boson}

The analysis for the non-compact boson is similar. For the Neumann case it follows from the analysis in Appendix~\ref{appCircle} that there is only one Ishibashi state (namely the one corresponding to $n=0$). As a consequence, there is no Wilson line, and the open string spectrum is simply given by $\eta^{-1}(\tau)$. We also note that the normalisation in eq.~(\ref{Dirbrane}) is formally infinite for $R\rightarrow \infty$, see also the discussion below eq.~(\ref{eq:IshibashiAdS2}).

On the other hand, if we impose a Dirichlet boundary condition, see eq.~(\ref{eq:gluingseed}), 
\be
(\alpha^0_n - \bar{\alpha}^0_{-n})|p\rangle\!\rangle = 0 \ ,
\label{eq:Is_R0}
\ee
the corresponding boundary states are given by
\be
|\!| x \rangle\!\rangle =  \int_{-\infty}^{\infty} dp \, e^{2\pi i p x} |p \rangle\!\rangle \ .
\label{eq:BS_R0}
\ee
The overlap of the Ishibashi states in eq.~(\ref{eq:Is_R0}) is
\begin{equation}\label{A.10}
\langle\!\langle {{p}}|\, e^{\pi i \hat{t}(L_0+\bar{L}_0-\frac{c}{12})}\,
| {{p}}\rangle\!\rangle = \chi_{{p}} ({\hat{t}}\,)  \ , \qquad \hbox{where} \qquad 
\chi_{{p}} ({\hat{t}}\,) =  \, \frac{e^{\pi i \hat{t} \, p^2}}{\eta(\hat{t}\,)} \ .
\end{equation}
Its $S$-modular transformation equals
\begin{equation}
   \chi_{{p}} \left(-\tfrac{1}{{t}}\right) = \int_{-\infty}^\infty ds' e^{-2\pi i ps'} \chi_{{s'}} (t) \ ,
\end{equation}
and hence the overlap of the boundary states in eq.~(\ref{eq:BS_R0}) leads to 
\begin{align}
\langle\!\langle y|\!|\, e^{\pi i \hat{t}(L_0+\bar{L}_0-\frac{c}{12})}\,
    |\!| x\rangle\!\rangle & = \int_{-\infty}^\infty d{p}\, e^{2\pi i {p}(x-y)} \, \chi_{p} ({\hat{t}}\,) \\
    & = \int_{-\infty}^\infty d{p}\, e^{2\pi i {p}(x-y)} \, \int_{-\infty}^\infty ds' e^{-2\pi i \,ps'} \chi_{s'} ({{t}}) = \, \chi_{(x-y)} ({{t}}) \ ,
\end{align}
where we have performed the usual transformation to the open string variable $t = -\tfrac{1}{\hat{t}}$. This then describes an open string stretching between $x$ and $y$.

\subsection{The fermionic boundary states}\label{appFermions}

For each individual fermion we impose the gluing condition 
\begin{align}
& (\psi_r -i \varepsilon \bar{\psi}_{-r}) | {\rm N},\varepsilon \,\rangle\!\rangle = 0 \\
& (\psi_r +i \varepsilon \bar{\psi}_{-r})  | {\rm D} ,\varepsilon\,\rangle\!\rangle = 0 \ , 
\end{align}
where `N' and `D' stands for `Neumann' and `Dirichlet' respectively, and $\varepsilon$ denotes the spin structure. Furthermore, in the NS-NS sector, $r$ runs over all half-integers, while in the R-R sector, $r$ runs over all integers. In each case, this fixes the Ishibashi state up to normalisation, and the normalisation is chosen so that 
\begin{align}
\langle \!\langle {\rm N},\varepsilon | e^{{\pi i \hat{\tau}(L_0+\bar{L}_0-\frac{1}{12})}} | {\rm N},\varepsilon \rangle\!\rangle_{\rm NS} & = \langle \!\langle {\rm D},\varepsilon| e^{{\pi i \hat{\tau}(L_0+\bar{L}_0-\frac{1}{12})}} | {\rm D},\varepsilon \rangle\!\rangle_{\rm NS} 
= \Bigl(\frac{\vartheta_3(\hat{\tau})}{\eta(\hat{\tau})}\Bigr)^{\frac{1}{2}} \ , \\ 
\langle \!\langle {\rm N},\varepsilon| e^{{\pi i \hat{\tau}(L_0+\bar{L}_0-\frac{1}{12})}} | {\rm N},\varepsilon \rangle\!\rangle_{\rm R} & = \langle \!\langle {\rm D},\varepsilon| e^{{\pi i \hat{\tau}(L_0+\bar{L}_0-\frac{1}{12})}} | {\rm D} ,\varepsilon\rangle\!\rangle_{\rm R} 
= \Bigl(\frac{\vartheta_2(\hat{\tau})}{\eta(\hat{\tau})}\Bigr)^{\frac{1}{2}} \ , \\ 
\langle \!\langle {\rm N},\varepsilon| e^{{\pi i \hat{\tau}(L_0+\bar{L}_0-\frac{1}{12})}} | {\rm D},\varepsilon \rangle\!\rangle_{\rm NS} & = \langle \!\langle {\rm D},\varepsilon| e^{{\pi i \hat{\tau}(L_0+\bar{L}_0-\frac{1}{12})}} | {\rm N} ,\varepsilon\rangle\!\rangle_{\rm NS} 
= \Bigl(\frac{\vartheta_4(\hat{\tau})}{\eta(\hat{\tau})}\Bigr)^{\frac{1}{2}} \ , \\ 
\langle \!\langle {\rm N},\varepsilon| e^{{\pi i \hat{\tau}(L_0+\bar{L}_0-\frac{1}{12})}} | {\rm D} ,\varepsilon\rangle\!\rangle_{\rm R} & = \langle \!\langle {\rm D},\varepsilon| e^{{\pi i \hat{\tau}(L_0+\bar{L}_0-\frac{1}{12})}} | {\rm N} ,\varepsilon\rangle\!\rangle_{\rm R} 
= \Bigl(\frac{\vartheta_1(\hat{\tau})}{\eta(\hat{\tau})}\Bigr)^{\frac{1}{2}} \ .
\end{align}
The overlaps for the cases where the two $\varepsilon$ do not agree follow by replacing for one state ${\rm N} \leftrightarrow {\rm D}$. 

\subsubsection*{The worldsheet case}
On the worldsheet all the fermions are simultaneously in the NS sector or the R sector, and we need to impose the GSO projection. This requires that we add (or subtract) the components with different values of $\varepsilon$. Finally, the actual worldsheet boundary states will have both an NS and an R component, see eq.~(\ref{eq:GSOstate}).

The overlaps for the boundary states involving all  ten fermions as in eq.~(\ref{eq:GSOstateNS}) and (\ref{eq:GSOstateR}) are then
\begin{subequations}\label{eq:NSRoverlaps}
\begin{align}
\hat{\chi}^+_{\rm NSNS} (\hat{\tau})= \,_{\text{S}}\langle \!\langle {\rm NS},\pm |\!| e^{{\pi i \hat{\tau}(L_0+\bar{L}_0-\frac{1}{12})}} \| {\rm NS},\pm \rangle\!\rangle_{\text{S}} & = \Bigl(\frac{\vartheta_3(\hat{\tau})}{\eta(\hat{\tau})}\Bigr)^{5} \ , \\
\hat{\chi}^-_{\rm NSNS} (\hat{\tau})= \,_{\text{S}}\langle \!\langle {\rm NS},\mp\| e^{{\pi i \hat{\tau}(L_0+\bar{L}_0-\frac{1}{12})}} \| {\rm NS},\pm \rangle\!\rangle_{\text{S}} & = \Bigl(\frac{\vartheta_4(\hat{\tau})}{\eta(\hat{\tau})}\Bigr)^{5} \ , \\
\hat{\chi}^+_{\rm RR}(\hat{\tau}) = \,_{\text{S}}\langle \!\langle {\rm R},\pm\| e^{{\pi i \hat{\tau}(L_0+\bar{L}_0-\frac{1}{12})}} \| {\rm R},\pm \rangle\!\rangle_{\text{S}} & = \Bigl(\frac{\vartheta_2(\hat{\tau})}{\eta(\hat{\tau})}\Bigr)^{5} \ , \\ 
 \hat{\chi}^-_{\rm RR}(\hat{\tau}) = \,_{\text{S}}\langle \!\langle {\rm R},\mp\| e^{{\pi i \hat{\tau}(L_0+\bar{L}_0-\frac{1}{12})}} \| {\rm R},\pm \rangle\!\rangle_{\text{S}} & = \Bigl(\frac{\vartheta_1(\hat{\tau})}{\eta(\hat{\tau})}\Bigr)^{5}=0 \ ,
\end{align}
\end{subequations}
where we have only considered self-overlaps, i.e.\ all fermions either satisfy DD or NN boundary conditions. The last overlap, i.e.\ $\hat{\chi}^-_{\rm RR}(\hat{\tau})$, is zero due to $\vartheta_1(\hat{\tau})=0$.

\subsubsection*{The dual CFT case}

For the dual CFT all the $8$ fermions of the seed theory are free fermions in the NS sector, and there is no GSO projection. As a consequence, their boundary state overlap is given by (see eq.~(\ref{eq:NSRoverlaps}))
\be
\langle \!\langle \pm | e^{{\pi i \hat{\tau}(L_0+\bar{L}_0-\frac{1}{12})}} | \pm \rangle\!\rangle  = \Bigl(\frac{\vartheta_3(\hat{\tau})}{\eta(\hat{\tau})}\Bigr)^{4} \ .
\ee

\section{Theta and eta functions}\label{app:theta}

In this paper we work with the conventions of e.g.\ \cite{Blumenhagen:2013fgp} for the theta functions:
\begin{align}
\vartheta\!\begin{bmatrix}
  \alpha \\
  \beta
  \end{bmatrix}(t;\tau)&\equiv\sum_{n\in\mathds Z}\mathrm{e}^{\pi i(n+\alpha)^2\tau+2\pi i(n+\alpha)\beta}\mathrm{e}^{2\pi i(n+\alpha)t}\\
  &=x^{\alpha}\mathrm{e}^{2\pi i\alpha \beta} q^{\frac{\alpha^2}{2}} \prod_{n=1}^\infty \big(1-q^n \big) \big(1+xq^{n+\alpha-\frac{1}{2}} \mathrm{e}^{2\pi i\beta} \big)\big(1+x^{-1}q^{n-\alpha-\frac{1}{2}} \mathrm{e}^{-2\pi i\beta} \big)\ ,
\end{align}
with $q = e^{2\pi i \tau}$ and $x=e^{2\pi i t}$.
The four Jacobi theta functions are given by
\be
\vartheta_1\equiv\vartheta\!\begin{bmatrix}
  \frac{1}{2} \\
  \frac{1}{2}
  \end{bmatrix}\ ,\qquad
\vartheta_2\equiv\vartheta\!\begin{bmatrix}
  \frac{1}{2} \\
  0
  \end{bmatrix}\ ,\qquad
\vartheta_3\equiv\vartheta\!\begin{bmatrix}
  0 \\
  0
  \end{bmatrix}\ ,\qquad
\vartheta_4\equiv\vartheta\!\begin{bmatrix}
  0 \\
  \frac{1}{2}
  \end{bmatrix}\ .
\ee
For the Dedekind $\eta$ function we have
\be
\eta(\tau) = e^{\frac{\pi i}{12} \tau} \, \prod_{n=1}^{\infty} (1 - e^{2\pi i \tau n}) \ .
\ee
We also recall that these functions satisfy the modular transformation properties
\be
\eta\left(-\tfrac{1}{\tau}\right) = \sqrt{-i\tau} \,\eta(\tau)  \ , \quad \vartheta_3(-1/\tau) =\sqrt{-i\tau} \,  \vartheta_3(\tau)\ , \quad  
\vartheta_4(-1/\tau) =\sqrt{-i\tau} \,  \vartheta_2(\tau)  \ .
\ee
Furthermore, we shall make use of the abstruse identity 
\be
\label{eq:abstruse}
\vartheta_3(\tau)^4 =  \vartheta_2(\tau)^4+\vartheta_4(\tau)^4 \ .
\ee

\bibliographystyle{JHEP}
\end{document}